\documentclass[12pt]{JHEP3}
\usepackage{amsfonts} 
\usepackage{amssymb}
\usepackage{lscape}
\usepackage{cite}
\usepackage{epsfig}
\usepackage{multirow}
\usepackage{longtable}
\usepackage{amsmath}

\usepackage{slashed}

\author{}

\def\a{\alpha}

\def\sla{\raise.15ex\hbox{$/$}\kern-.57em}
\let\a=\alpha         \let\d=\delta
\let\e=\epsilon         
    \let\k=\kappa    \let\m=\mu
\let\n=\nu

\newcommand{\be}{\begin{equation}}
\newcommand{\ee}{\end{equation}}
\newcommand{\ba}{\begin{array}}
\newcommand{\ea}{\end{array}}
\newcommand{\bea}{\begin{eqnarray}}
\newcommand{\eea}{\end{eqnarray}}

\newcommand{\ov}{\overline}
\def\IR{\relax{\rm I\kern-.18em R}}

\def\IP{\relax{\rm I\kern-.18em P}}
\def\inbar{\vrule height1.5ex width.4pt depth0pt}
\def\IC{\relax\,\hbox{$\inbar\kern-.3em{\rm C}$}}

\def\a{\alpha}

\def\K3{{\bf K3}}

\def\d{\delta}

\def\ov{\overline}

\def\n2d{\cN_{V^*}^{\otimes 2}}

\def\IC{\mathbb{C}}

\def\IR{\mathbb{R}}

\def\IP{\mathbb{P}}

\def\cN{{\mathcal N}}

\def\nn{\nonumber}

\def\Tr{{\mathop {\rm Tr}}}
\def\to{\rightarrow}

\title{Production of light stringy states}
\author{
Pascal Anastasopoulos$^{1}$\footnote{pascal@hep.itp.tuwien.ac.at} and
Robert Richter$^{2}$\footnote{robert.richter@desy.de}~
\\
$^1$ Technische Univ. Wien Inst. f\"ur Theoretische Physik, A-1040 Vienna, Austria\\
$^2$ II. Institut f\"ur Theoretische Physik, Hamburg University, Germany\\
}
\date{}

\maketitle
\abstract{We discuss the direct production of light stringy states. Those states correspond to the lightest stringy excitations localized at the intersection of two D-brane stacks and their  mass can be parametrically smaller than the string scale. Thus in a low string scale scenario one may observe stringy signatures at LHC even in the absence of the usual string
resonances. We compute the tree level string scattering amplitude involving two gauge bosons and two such light stringy states. We give the squared amplitude summed over all polarization and gauge configuration and eventually apply the result to the quark sector of a D-brane realization of the Standard Model. 
}
\preprint{TUW-14-11\\ 
ZMP-HH/14-17}
 \clearpage

\begin{document}

\section{Introduction}

D-brane compactifications have been proven to be a successful framework for realistic model building \footnote{For recent reviews on D-brane model building as well as specific MSSM D-brane constructions, see \cite{Blumenhagen:2005mu, Blumenhagen:2006ci, Marchesano:2007de, Cvetic:2011vz} and references therein.}. In those string constructions the gauge bosons are strings attached to a stack of lower-dimensional hyperplanes, so called D-branes, while the chiral matter, such as the Standard Model (SM) fermions, are strings that appear at the intersection of different D-brane stacks. 

In such constructions the four-dimensional Planck mass depends on the whole  internal volume while the gauge couplings of the gauge theory living on a D-brane stack is controlled by the volume of the cycle the D-brane wraps in the compactified dimension. Thus compactifications which exhibit a hierarchy in the volumes of the cycles wrapped by the SM D-brane sector (\emph{small cycles}) and the transverse dimensions (\emph{large dimensions}) give rise to a small string scale. In principle after moduli stabilization one may even end up with a Calabi-Yau manifold that large, that the string scale $M_s$ is of the order of a few TeV \cite{ArkaniHamed:1998rs, Antoniadis:1997zg, Antoniadis:1998ig}. Such a scenario   provides a solution to the hierarchy and cosmological constant problems but may also lead to interesting signatures observable at the LHC such as signs of anomalous $Z'$ physics as well as Kaluza Klein states (see, e.g.\cite{Dudas:1999gz, Accomando:1999sj, Cullen:2000ef, Kiritsis:2002aj, Antoniadis:2002cs, Ghilencea:2002da, Anastasopoulos:
2003aj, Anastasopoulos:2004ga,  Burgess:2004yq, Burikham:2004su, Chialva:2005gt, Coriano':2005js,  Bianchi:2006nf, Anastasopoulos:2006cz, Anchordoqui:2007da,   Anastasopoulos:2008jt, Anchordoqui:2008ac, Lust:2008qc ,Anchordoqui:2008di, 
Armillis:2008vp, Fucito:2008ai, Anchordoqui:2009mm, Anchordoqui:2009ja, Lust:2009pz, Dong:2010jt, Feng:2010yx, Anchordoqui:2010zs,Cicoli:2011yy, Anchordoqui:2012wt, Anchordoqui:2011eg, Anchordoqui:2011ag, Hashi:2012ka, Lust:2013koa, Anchordoqui:2014wha})\footnote{For a recent review, see \cite{Berenstein:2014wva}.}.

\begin{figure}[h]
\begin{center}
\epsfig{file=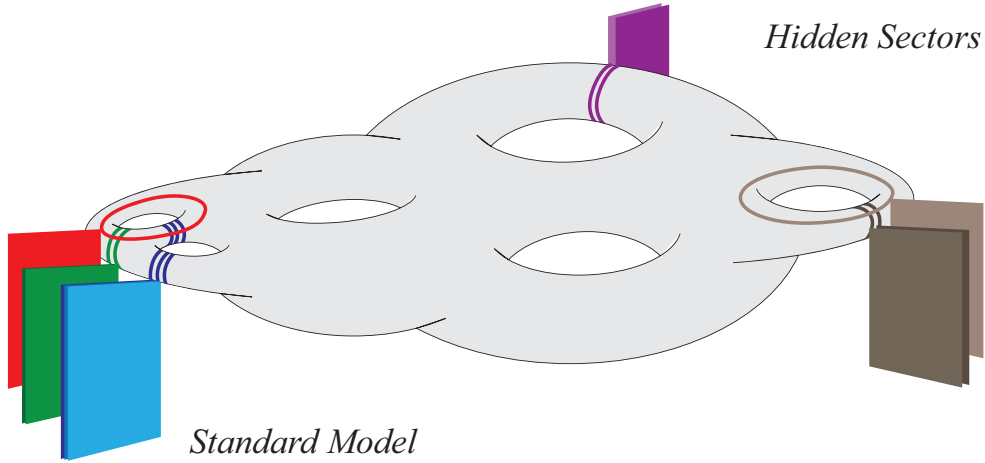,width=130mm}
\caption{Low string scale scenario.
}\label{fig. low string scale CY}
\end{center}
\end{figure}
Such a scenario is displayed in figure \ref{fig. low string scale CY}, where the Standard Model sector is localized on a small region of the full Calabi-Yau manifold. Note that the cycles wrapped by the SM D-brane sector are small, thus reproducing the observed SM gauge couplings, while the overall volume of the Calabi-Yau is large, allowing for a string scale in the TeV region opening the possibility for the direct detection of string excitations. 

In a series of papers \cite{Lust:2008qc,Anchordoqui:2009mm,Lust:2009pz,Anchordoqui:2009ja} the authors study tree-level string scattering amplitudes containing massless bosons and fermions that can be identified with the SM fields. They show that amplitudes containing at most two chiral fermions exhibit a universal behaviour independently of the specifics of the compactification. The  poles of these scattering amplitudes correspond to exchanges of Regge excitations of the SM gauge bosons, whose mass scales like the string mass. Due to the universal behaviour of these amplitudes, which in turn give them a predictive power, LHC is able to constrain the string scale to be above $4.5$ TeV
\cite{Chatrchyan:2011ns, ATLAS:2012pu,  Chatrchyan:2013qha}.

\begin{figure}[h]
\begin{center}
\epsfig{file=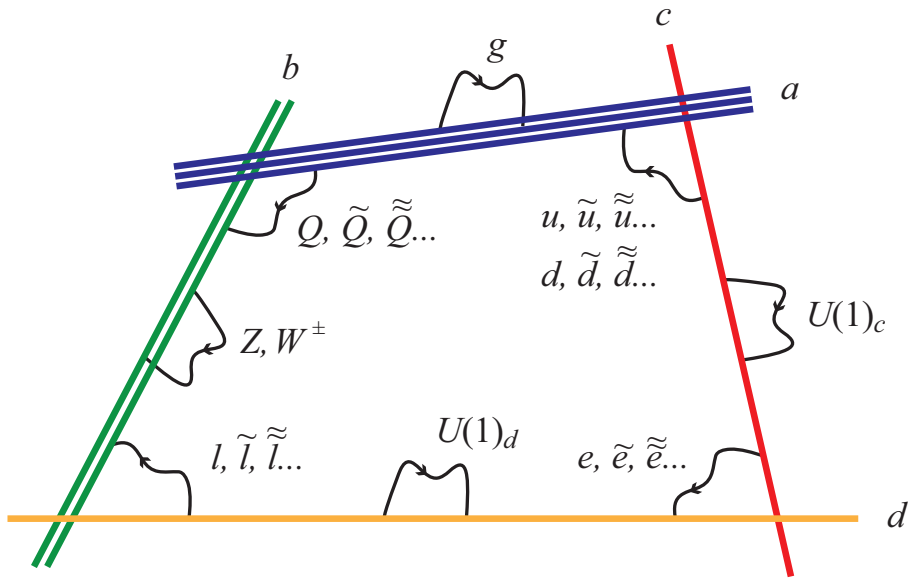,width=120mm}
\caption{Local D-brane realization of the Standard Model.
}\label{fig. madrid}
\end{center}
\end{figure}

At the intersection of two D-brane stacks there exists beyond the massless fermion a whole tower of stringy excitations, so called \emph{light stringy states}, whose mass is $M^2 =\theta M^2_s$, where $\theta$ denotes the intersection angle between these two D-brane stacks \cite{Anastasopoulos:2011hj}. Given a small intersection angle those light stringy states can be significantly lighter than the first Regge excitations of the gauge bosons and are expected to be observed prior to the latter. In figure \ref{fig. madrid} we depict a D-brane SM realization.  While the SM gauge bosons live on the world volume of the D-brane stacks the SM fermions are localized at the intersections of different D-brane stacks. At each intersection there exist a  tower of massive stringy excitations that can be significantly lighter than the string scale for some regions in the parameter space. Let us stress that in this setup one has for each massless fermion a separate tower of stringy excitations with a different mass spacing. In that respect it can be easily discriminated from Kaluza-Klein scenarios, in which generically the mass gap is for each particle the same.

In this paper we plan to further investigate those light stringy states. More precisely, we compute 4-point tree level string scattering amplitudes containing two gauge bosons and two light stringy states. A crucial ingredient in this calculation is the exact knowledge of the vertex operators of the light stringy states, which we derive applying the dictionary laid out in \cite{Anastasopoulos:2011hj}. Equipped with those we then compute the full string scattering amplitude and sum over all polarization and gauge configurations of the squared amplitude. Eventually we apply our results to a D-brane realization of the Standard Model, thus bringing it into a form suitable to comparison to LHC data.

The paper is organized as follows. In section \ref{sec vertex operator} we discuss in detail the vertex operators of light stringy states. Given those in section \ref{sec scattering amplitude} we evaluate the tree level string scattering amplitude involving two gauge bosons and two light stringy states. In section \ref{sec cross section} we compute the squared amplitudes summed over all the polarization and gauge configurations of initial and final particles involved. Eventually we apply our findings to a D-brane realization of the Standard Model.

\section{Vertex operators \label{sec vertex operator}}

Before discussing the vertex operators of light stringy states let us briefly describe the setup in which we perform our analysis \footnote{For more details, see \cite{Lust:2008qc}, where the authors use an analogous setup.}. Our explicit string computation will be carried out for type IIA, however analogous results hold true for the T-dual type IIB picture. To be more precise, in order to ensure calculability of the string amplitudes, we assume that the region in the Calabi-Yau manifold where SM D-brane sector is localized, does locally look like a factorizable six-torus $T^6=T^2 \times T^2 \times T^2$, analogously to the assumption in \cite{Lust:2008qc}. Then the D6-branes will wrap 3 cycles within this local factorizable six-torus.   
\begin{figure}[h]
\begin{center}
\epsfig{file=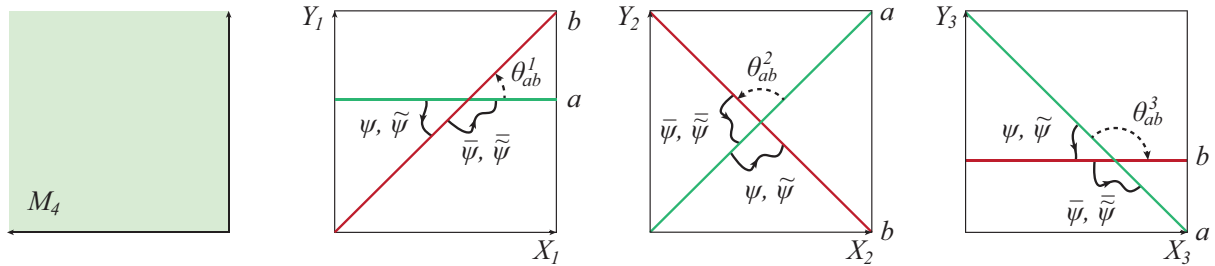,width=160mm}
\caption{Two intersecting D-brane stacks on a $T^2 \times T^2 \times T^2$.
}\label{fig. intersecting branes}
\end{center}
\end{figure}

In figure \ref{fig. intersecting branes} we display an intersection between such two D-brane stacks on a factorizable six-torus. As pointed out in \cite{Anastasopoulos:2011hj} beyond the massless fermion $\psi$ \footnote{To keep the discussion general we denote with $\widetilde{\psi}$ a massive light stringy state appearing at an arbitrary intersection of two D-brane stacks. In the context of the SM $\widetilde{\psi}$ can be the lightest stringy excitation of any massless fermionic SM field, as depicted in figure \ref{fig. madrid}.} 
there exists a whole tower of stringy excitations \footnote{Note that for each two-torus there exists a set of independent creation operators, which can be applied to the ground state.  For details, see \cite{Anastasopoulos:2011hj}.}, with the lightest being $\widetilde{\psi}$. The mass depends on the smallest intersection angle, which without loss of generalization we assume from now on  to be the intersection angle in the first two-torus, $\theta^1_{ab}$. 

For the proper normalization of the vertex operators of the gauge bosons as well as of the massive fermionic states $\widetilde{\psi}$, discussed in detail in the subsequent section, let us display the low energy effective action for those particles
\begin{align}
{\cal L} = - Tr \sum_{a} \frac{1}{2\, g^2_{Dp_a}} F^a_{\mu\nu}F^{a\,\, \mu \nu}  - Tr \sum_{a \cap b} \left( i \ov \psi \, \slashed{D}^{a \cap b} \,\psi +  
 i {\ov {\widetilde \psi} } \,\widetilde{\slashed{  D}}^{a \cap b}\,
  {\widetilde \psi}
  \right) + m \,  {\ov {\widetilde \psi}} {{\widetilde \psi} } + \ldots \,\,.
 \label{eq effective action}
\end{align}
Here $F^a_{\mu \nu}$ denotes the field strength of the gauge boson $A^a$, while $\psi$ and $\widetilde{\psi}$ denote the massless and massive fermion localized at the intersection of the two D-brane stacks $a$ and $b$. The covariant derivative ${D}^{a \cap b}_{\mu}$ takes the form
\begin{align}
{D}^{a \cap b}_{\mu}  = \partial^{\mu}    +  T^a A^a_{\mu} - i T^b A^b_{\mu}
\end{align}
with $T^a$ and $T^b$ being the generators of the respective gauge symmetries. Finally the ellipses indicate additional terms, such as the kinetic terms for the super partners as well as interaction terms.   

The aim of this work is to compute the string four-point scattering amplitude of two gauge bosons into the lightest stringy excitation of the massless chiral fermion localized at the intersection of two D6-branes. More precisely, we want to calculate the string amplitude
\begin{align}
{\cal M}=\int \frac{\prod^4_ {i=1} dz_i }{ V_{CKG}}
  \Big\langle V^{(0)}_A [z_1, \epsilon_1, k_1] \,  V^{(-1)}_A [z_2, \epsilon_2, k_2] \,V^{(-1/2)}_{\widetilde{\psi}} [z_3 , v_3, {\overline u}_3, k_3 ] 
 \,V^{(-1/2)}_{\ov {\widetilde \psi }} [z_4 , {\overline v}_4, u_4, k_4 ] \Big\rangle \,\,,
\label{eq amplitude}
\end{align}
where $A$ denotes the gauge field living on the D-brane, ${\widetilde \psi}$  denotes the lightest stringy excitation of the massless chiral fermion localized at the intersection of two D6-brane stacks $a$ and $b$, and $\ov { \widetilde \psi}$ being the complex conjugated counterpart of $\widetilde{\psi}$. In the following we discuss and display the vertex operators of the respective particles that are required to carry out the computation of \eqref{eq amplitude}.

Let us start by displaying the vertex operators for the gauge bosons. In the (-1)-ghost picture it takes the form 
\begin{align} 
V^{(-1)}_A [z, \epsilon, k] = g_A [T^a]^{\alpha_1}_{\alpha_2} \, e^{-\phi} \epsilon^{\mu} \,  \psi_{\mu} e^{i k X}    
\label{eq gauge boson -1 ghost}
\end{align}
while in the (0)-ghost picture it is given by
\begin{align} 
V^{(0)}_A [z, \epsilon, k] = \frac{g_A}{\sqrt{2 \alpha'}} [T^a]^{\alpha_1}_{\alpha_2} \,  \epsilon^{\mu} \Big\{ \partial X_{\mu} - 2 i \alpha' (k \cdot \psi) \,  \psi_{\mu}\Big\} e^{i k X}    \,\,.
\label{eq gauge boson 0 ghost}
\end{align}
Here $\epsilon^{\mu}$ is the polarization vector, while $[T^a]^{\alpha_1}_{\alpha_2}$ denotes the Chan-Paton factor, where $a$ denotes the D-brane stack on which the gauge boson is localized and the indices $\alpha_1$ and $\alpha_2$ describe the two string ends. Finally, the string vertex coupling $g_A$ can be determined by computing the three gauge boson disk scattering amplitude and comparing it with the low energy effective action \eqref{eq effective action} yielding to \footnote{Here we use the normalization factor  $C_{D2}= \alpha'^{-2} \,g^{-2}_{Dp_a} $ for disk amplitudes containing strings localized solely on a single D-brane stack \cite{Polchinski:1998rq, Polchinski:1998rr, Lust:2008qc}.}
\begin{align}
g_A= \sqrt{2 \alpha'} \, g_{Dp_a}\,\,.
\end{align}
 
Now let us turn to the fermionic vertex operator localized at the intersections of two D-brane stacks $a$ and $b$. We assume that the two D-brane stacks intersect such on the $T^2 \times T^2  \times T^2 $ that the three intersection angles are $\theta^1_{ab}>0$, $\theta^2_{ab}>0$ and  $\theta^3_{ab}<0$, see figure \ref{fig. intersecting branes}. Though it is irrelevant for the analysis performed here we assume that the two D-brane stacks intersect such that ${\cal N}=1$ supersymmetry is preserved, which for the setup at hand implies 
\begin{align}
\theta^1_{ab} + \theta^2_{ab} + \theta^3_{ab} = 0 \,\,.
\label{eq SUSY condition}
\end{align}
Here the angles are measured in multiples of $\pi$.

The R-sector contains a massless fermion given by the R-vacuum $| \, \theta \,\rangle_R$, whose vertex operator for the concrete choice of intersection angles \eqref{eq SUSY condition} takes the form \footnote{For a detailed discussion on vertex operators of massless states for arbitrary intersection angles,
see \cite{Cvetic:2006iz,Bertolini:2005qh}, for a generalization to massive states see \cite{Anastasopoulos:2011hj} and for a discussion on instantonic states
at the intersection of D-instanton and D-brane at arbitrary angles, see \cite{Cvetic:2009mt}.}
\begin{align}
  | \, \theta \,\rangle^{ab}_R : ~~~  V^{(-1/2)}_{\psi} = g_{\psi}\,  [T^{ab}]^{\beta_1}_{\alpha_1} \, e^{-\varphi/2}   \,v^{\alpha} \,S_{\alpha} \,\,\prod^2_{I=1} \sigma_{\theta^I_{ab}} \, e^{i\left(\theta^I_{ab}- \frac{1}{2} \right)H_I}  
  \sigma_{1+\theta^3_{ab}} \, e^{i\left(\theta^3_{ab}+ \frac{1}{2} \right)H_3}
  \,\, e^{ikX} \,\,,
  \label{eq vertex operator massless}
\end{align}
where the Chan-Paton factors indicate that the string is stretching between the two D-brane stacks $a$ and $b$. Here the indices $\alpha_1$ and $\beta_1$ run from one to the dimension of the fundamental representation of the gauge group living on the respective D-brane stack $a$ and $b$. The internal part of the vertex operators contains bosonic and fermionic twist fields $\sigma_{\alpha_I}$ and $e^{-i \alpha_I H_I}$ introduced to account for the mixed boundary conditions of the open string stretched between the two intersecting D-brane stacks. The GSO projection determines the chirality of the four-dimensional polarization spinor. It can be easily identified by analyzing the $U(1)$ world-sheet charge of the respective vertex operator. The latter is given by the sum $\sum^3_{I=1} q_I$, where the $q_I$ are the prefactors appearing in front of the  bosonized internal fermionic fields, i.e. in front of the $H_I$. If the sum adds up to $-\frac{1}{2} \mod 2$ the spin polarization is chiral, while if the sum is $\frac{1}{2} \mod 2$ the polarization is anti-chiral. 

Let us check the BRST-invariance of the vertex operator, where the BRST- charge takes the form
\begin{align}
Q_{BRST}&= \oint \frac{dz}{2 \pi i}\bigg\{ e^{\varphi} \eta  \frac{1}{\sqrt{2 \alpha'}} \bigg( i\partial X^{\mu} \, \psi_{\mu} + \sum^3_{I=1} \partial Z^{I } \, e^{-iH_I} + \sum^3_{I=1} \partial \ov Z^{I } \, e^{iH_I}\bigg) \label{eq BRST charge}\\
&~~~~+ c \bigg(\frac{1}{\a'} i\partial X^\mu i \partial X_\mu  - \frac{1}{2} \psi^\mu \partial \psi_\mu + \sum^3_{I=1} \Big( \frac{1}{\a'} \partial Z^I \partial \ov Z_I - \frac{1}{2} e^{-iH_I} \partial e^{iH_I} \Big)\bigg) \bigg\} + ...\nn
\end{align}
with the ellipses indicate additional terms containing purely ghost contributions which are irrelevant for our analysis. Here we introduced the complex coordinates $\partial Z^I = \frac{1}{\sqrt{2}}\left(\partial X^{I} + i \partial Y^{I}\right)$ with $I=1,2,3$ denoting the internal (compactified) dimensions. Given \eqref{eq BRST charge} one obtains, using the operator product expansions (OPE's ) displayed in appendix \ref{app OPE}, that BRST invariance is ensured if one requires 
\begin{align}
\alpha' k^2=0 \qquad \qquad   v^{\alpha}  \sqrt{\alpha'}    k^{\mu}  \sigma_{\alpha \dot{\alpha}} = 0
\label{eq EOM massless}  \,\,,
\end{align}
where  the second part of \eqref{eq EOM massless} is the EOM for a massless Weyl fermion. Finally, the string vertex coupling $g_{\psi}$ can be derived by calculating the three point amplitude $\langle A \psi \ov \psi \rangle $ and comparing it to the effective action \eqref{eq effective action} in an analogous fashion as before for the gauge boson vertex coupling $g_A$. One obtains 
\begin{align}
g_{\psi} = \sqrt{ 2}  \alpha'^{\frac{3}{4}} e^{\phi_{10}/2} \,\,,
\end{align}
where we used the normalization factor for a disk scattering amplitude containing strings stretched between different D-brane stacks that is given by  $\widetilde{C}_{D2}= \alpha'^{-2} e^{-\phi_{10}}$.

As pointed out above there are stringy excitations which in case of a low string scale may be observable at LHC in the near future. Let us focus on lightest string excitations, which assuming that $\theta^1_{ab}$ is the smallest of all three intersection angles \cite{Anastasopoulos:2011hj} are given by
\begin{align}
\alpha_{-\theta^1_{ab}}    | \, \theta \,\rangle^{ab}_R
 \qquad \qquad \text{and} \qquad \qquad \psi_{-\theta^1_{ab}}   | \, \theta \,\rangle^{ab}_R\,\,.
\end{align}
Here $| \, \theta \,\rangle^{ab}_R$ denotes the ground state of the R-sector, which corresponds to the massless fermion and $\alpha_{-\theta^1_{ab}} $ and $ \psi_{-\theta^1_{ab}} $ are the bosonic and fermionic creation operators. Both states have the same mass, $\alpha' m^2 = \theta^1_{ab}$, and combine into a massive Dirac spinor $\widetilde\psi$, whose vertex operator takes the form using the dictionary derived in \cite{Anastasopoulos:2011hj} and summarized in appendix \ref{app dictionary}
\begin{align}
V^{(-1/2)}_{\widetilde \psi} = & g_{\widetilde{\psi}}\,[T^{ab}]^{\beta_1}_{\alpha_1} e^{-\varphi/2} \left( \frac{\widetilde v_{\a}}{\sqrt{\theta^1_{ab} }} \,S^{\a} \,{\tau}_{\theta^1_{ab} }\, e^{i(\theta^1_{ab} -\frac{1}{2}) H_1} + \ov {\widetilde u}_{\dot{\a}} \,S^{\dot{\a}}\, \sigma_{\theta^1_{ab} }\, e^{i(\theta^1_{ab}  +\frac{1}{2}) H_1}  \right)  \nn\\
& \hspace{40mm} \times \sigma_{\theta^2_{ab} } e^{i(\theta^2_{ab}  -\frac{1}{2}) H_2} 
\, \sigma_{1+\theta^3_{ab} } e^{i(\theta^3_{ab}  +\frac{1}{2}) H_3}
\,e^{ikX}\,\,.
\label{eq. vertex operator massive fermion}
\end{align}
Here $\widetilde{v}$ and $\ov {\widetilde u}$ denote the left and right-moving Weyl fermions of the light stringy state $\widetilde{\psi} = (\widetilde{v}_{\a}, \ov { \widetilde u}_{\dot{\a}})$, where the respective chirality is dictated by the $U(1)$ world-sheet charge. Note that in contrast to the vertex operator of the massless fermion (see eq. \eqref{eq vertex operator massless}) the vertex operator of $\widetilde{\psi }$ contains not only regular twist fields but also the higher excited bosonic and fermionic twist fields. One can easily check that BRST invariance requires 
\begin{gather}
\alpha' k^2 = - \theta^1_{ab} \\
\widetilde v_{\a} \, \sqrt{\alpha'} \,k^{\mu} {\sigma_{\mu}}^{\a \dot{\a}} + \sqrt{\theta^1_{ab}} \ov {\widetilde u}_{\dot{\a}}=0 \qquad \qquad 
\ov {\widetilde u}_{\dot{\a}} \,\sqrt{\alpha'}\, k_{\mu} \ov \sigma^{\dot{\a} \a}_{\mu} + \sqrt{\theta^1_{ab}}\widetilde v_{\a}=0\,\,,
\label{eq. dirac equation BRST invariance}
\end{gather}
where the second line \eqref{eq. dirac equation BRST invariance} is the Dirac equation written in terms of the left and right-handed Weyl components of the massive Dirac spinor $\widetilde{\psi}$.

In order to compute the string amplitude \eqref{eq amplitude} we need in addition to the gauge boson vertex operators \eqref{eq gauge boson -1 ghost} and \eqref{eq gauge boson 0 ghost} and the fermionic vertex operator \eqref{eq. vertex operator massive fermion} also the vertex operator of the complex conjugated of the massive Dirac spinor $\widetilde{\psi}$. It is given by the state 
\begin{align}
\widetilde{\alpha}_{-\theta^1_{ab}}    | \, \theta \,\rangle^{ba}_R
 \qquad \qquad \text{and} \qquad \qquad \widetilde{\psi}_{-\theta^1_{ab}}   | \, \theta \,\rangle^{ba}_R\,\,,
\end{align}
where the operators $\widetilde \alpha$ and $\widetilde{\psi}$ do correspond to the creation operators of a system of intersecting D6-branes system with opposite intersection angles\footnote{Note that here we consider the D-brane intersection $ba$, which in contrast to the system $ab$ does have two negative angles in the first two complex internal dimensions and a positive intersection angle in the third complex internal dimension.}. The corresponding vertex operator takes the form    
\begin{align}
V^{(-1/2)}_{\ov {\widetilde \psi}} = & g_{\widetilde{\psi}}\,[T^{ba}]^{\alpha_1}_{\beta_1} e^{-\varphi/2} \left( \frac{ \ov {\widetilde v}_{\dot{\a}}}{\sqrt{\theta^1_{ab} }} \,S^{\dot{\a}} \, \widetilde\tau_{1-\theta^1_{ab} }\, e^{-i(\theta^1_{ab} -\frac{1}{2}) H_1} +  {\widetilde u}_{\a} \,S^{\a}\, \sigma_{1-\theta^1_{ab} }\, e^{-i(\theta^1_{ab}  +\frac{1}{2}) H_1}  \right)  \nn\\
& \hspace{40mm} \times \sigma_{1-\theta^2_{ab} } e^{-i(\theta^2_{ab}  -\frac{1}{2}) H_2} \,\,\sigma_{-\theta^3_{ab} } e^{-i(\theta^3_{ab} +\frac{1}{2}) H_3}  \,e^{ikX}\,\,.
\label{eq. vertex operator massive antifermion}
\end{align}
where BRST invariance implies for the left- and right-moving Weyl components $\widetilde u_{\a}$ and $\ov {\widetilde v}_{\dot{\a}}$ to satisfy the Dirac equation, analogously  to the Weyl components of massive fermion vertex operator \eqref{eq. vertex operator massive fermion}. 

Finally, in an analogous fashion as before for the $g_{\psi}$ we determine $g_{\widetilde{\psi}}$ by computing the three-point function $\langle A \widetilde{\psi} \ov {\widetilde \psi}  \rangle $ and comparing it to the effective action \eqref{eq effective action} resulting in 
\begin{align}
g_{\widetilde{\psi}} = \sqrt{ 2}  \alpha'^{\frac{3}{4}} e^{\phi_{10}/2} \,\,.
\end{align}

Given \eqref{eq gauge boson -1 ghost},  \eqref{eq gauge boson 0 ghost}, \eqref{eq. vertex operator massive fermion} and  \eqref{eq. vertex operator massive antifermion} we have now all the four vertex operators required to perform the computation of the disk scattering amplitude \eqref{eq amplitude}.

\section{The scattering amplitude \label{sec scattering amplitude}}

In the previous section we derived all the vertex operators necessary to compute the amplitude containing two gauge bosons the two massive fermions ${\widetilde \psi}$ 
and ${\ov {\widetilde \psi}}$. Using the correlators displayed in appendix \ref{app OPE} we obtain for  the scattering amplitude \eqref{eq amplitude} 
\begin{align}
{\cal M} \sim \sqrt{\alpha'} \,\widetilde{C}_{D2}\, g_{A_x}  g_{A_y}\,
g^2_{\widetilde{\psi}}  \int \frac{\prod^4_{i=1} d x_i}{ V_{CKG}} \,\,x^{s-1}_{12}  \,  x^{t+\theta^1_{ab}}_{13} \,  x^{u+\theta^1_{ab}-1}_{14}   
x^{u+\theta^1_{ab}-1}_{23} \,  x^{t+\theta^1_{ab}}_{24}  \, x^{s-1}_{34} ~~ \nn\\
\times \left\{ \left(  (k_2 \cdot \epsilon_1) {\epsilon_2}_{\nu}  - ( k_1 \cdot \epsilon_2) {\epsilon_1}_{\nu} + (\epsilon_1 \cdot \epsilon_2) {k_1}_{\nu}  + \frac{x_{12}\, x_{34}}{x_{13}\, x_{24}} (k_3 \cdot \epsilon_1){\epsilon_2}_{\nu} 
\right) 
\right. ~~~~\nn\\
\left.
\times \left( ( {\widetilde v}_3 \, \sigma^{\nu} {\ov {\widetilde v}}_4 ) + (\ov {\widetilde u}_3 \, \ov \sigma^{\nu} {\widetilde u}_4 ) \right)  \right.~~~~ \nn \\
\left. 
+\frac{1}{2} \frac{x_{12}\, x_{34}}{x_{13}\, x_{24}} \, {\epsilon_1}_{\mu}\,  {\epsilon_2}_{\nu} \, {k_1}_{\lambda} \left(  ({\widetilde v}_3 \, \sigma^{\lambda} \ov \sigma^{\mu} \sigma^{\nu} \, {\ov {\widetilde v}}_4 ) + (\ov {\widetilde u}_3 \, \ov \sigma^{\lambda}  \sigma^{\mu} \ov \sigma^{\nu} {\widetilde u}_4 ) \right)
\right\} ~.
\end{align}
Here $x_{ij}=x_i-x_j$ and we dropped for the moment the Chan-Paton factors which will determine the relative positions of the respective vertex operator. 
As we will see momentarily the order of the Chan Paton factors depends crucially on whether the two gauge bosons from the same, $x=y$ or from different D-brane stacks $x\neq y$. The Mandelstam variables $s$, $t$ and $u$ are given by the usual expressions
\begin{align}
s=\alpha' \left(k_1 +k_2 \right)^2
\qquad t=\alpha' \left(k_1 +k_3 \right)^2
\qquad u=\alpha' \left(k_1 +k_4 \right)^2
\end{align}
and satisfy the relation
\begin{align}
s+ t + u = - 2  \theta^1_{ab}\,\,.
\end{align}

Finally, the factor $V_{CKG}$ is the volume of the conformal Killing group of the disk, which can be accounted for by fixing three vertex operator positions and adding the appropriate c-ghost correlator. While the generic four point amplitude is given by the sum over all 6 cyclic invariant orderings of the vertex operators, where each ordering gives rise to different integration region, the trace over Chan-Paton factors of the respective vertex operators gives only non-vanishing amplitudes for specific integration regions. 

\begin{figure}[h]
\begin{center}
\epsfig{file=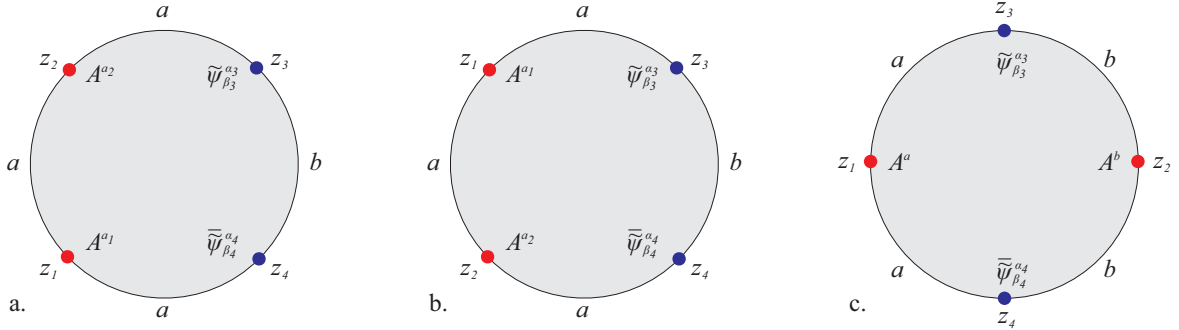,width=155mm}
\caption{The possible orderings of the respective vertex operators.
}\label{fig. disk positions}
\end{center}
\end{figure}

Fixing the vertex operator positions to be
\begin{align}
x_1=0 \qquad x_3 =1 \qquad x_4=\infty 
\end{align}
with both gauge bosons arising from the same D-brane stack (see figure \ref{fig. disk positions}.a and \ref{fig. disk positions}.b) we have the two integration regions $-\infty < x_2< 0$ 
and $0< x_2 <1$. On the other hand if the two gauge bosons arise from different D-brane stacks we have only one integration region, namely $1< x_2 < \infty$ (see figure \ref{fig. disk positions}.c). Adding the c-ghost correlator
\begin{align}
\Big\langle  c(x_1) c(x_3) c(x_4)\Big\rangle= x_{13} \, x_{14} \, x_{34} 
\end{align}
due to the fixing of the three vertex operator positions we obtain for the amplitude with the two gauge bosons living on the same D-brane stack
\begin{align}
&{\cal M}\left[ A^{a_1} [{\epsilon_1}, k_1], A^{a_2}[{\epsilon_2}, k_2], \widetilde{\psi}[{\widetilde v}_3, \ov {\widetilde v}_3, k_3 ], \ov {\widetilde \psi} [\ov {\widetilde v}_4, {\widetilde u}_4, k_4 ]  \right]=-2 \alpha' g^2_{Dp_a} \left( K + \widetilde{K}  \right) \nn \\
&~~~~~~~~~~~~~~~~\times \left\{ \Tr\left[T^{a_1} T^{a_2} {T^{ab}}{T^{ba}}\right] B[S, U] +  \Tr\left[T^{a_2} T^{a_1} T^{ab}T^{ba}\right] \frac{T}{U} B[S, T] \right\}
\end{align}
where the kinematic factors $K$ and  $\widetilde{K}$ take the form
\begin{align}
K&= \left\{ (k_2 \cdot \epsilon_1) {\epsilon_2}_{\nu}  - ( k_1 \cdot \epsilon_2) {\epsilon_1}_{\nu} + (\epsilon_1 \cdot \epsilon_2) {k_1}_{\nu}  - \frac{S}{T} (k_3 \cdot \epsilon_1){\epsilon_2}_{\nu} \right\} ( {\widetilde v}_3 \, \sigma^{\nu} {\ov {\widetilde v}}_4 )  \nn \\
& \hspace{30mm}-\frac{1}{2} \frac{S}{T} \, {\epsilon_1}_{\mu}\,  {\epsilon_2}_{\nu} \, {k_1}_{\lambda} \left(  {\widetilde v}_3 \, \sigma^{\lambda} \ov \sigma^{\mu} \sigma^{\nu} \, {\ov {\widetilde v}}_4 \right) \label{eq. K}\\
\widetilde{K}&=\left\{ (k_2 \cdot \epsilon_1) {\epsilon_2}_{\nu}  - ( k_1 \cdot \epsilon_2) {\epsilon_1}_{\nu} + (\epsilon_1 \cdot \epsilon_2) {k_1}_{\nu}  - \frac{S}{T} (k_3 \cdot \epsilon_1){\epsilon_2}_{\nu} \right\} \left(\ov {\widetilde u}_3 \, \ov \sigma^{\nu} {\widetilde u}_4 \right) \nn \\
& \hspace{30mm} -\frac{1}{2} \frac{S}{T} \, {\epsilon_1}_{\mu}\,  {\epsilon_2}_{\nu} \, {k_1}_{\lambda} \left(  \ov {\widetilde u}_3 \, \ov \sigma^{\lambda}  \sigma^{\mu} \ov \sigma^{\nu} {\widetilde u}_4  \right)\,\,.
\label{eq. tilde K}
\end{align}
Here $B[m,n]$ denotes the Euler Beta function that can be represented by the integral 
\begin{align}
B[m,n] = \int^1_0 x^{m-1} (1-x)^{n-1}= \frac{\Gamma[m] \, \Gamma[n]}{\Gamma[m+n]} ~.
\end{align}
For the sake of clarity we introduced the modified Mandelstam variables, defined as
\begin{align}
S=s
\qquad T=t+\theta^1_{ab}
\qquad U=u+\theta^1_{ab} 
\label{eq def of S T U}
\end{align}
that satisfy the relation
\begin{align}
S+ T + U = 0\,\,.
\end{align}

On the other hand for the two gauge bosons arising from two different stacks of D-branes we obtain
 \begin{align}
{\cal M}\left[ A^{a} [{\epsilon_1}, k_1], A^{b}[{\epsilon_2}, k_2], \widetilde{\psi}[{\widetilde v}_3, \ov {\widetilde v}_3, k_3 ], \ov {\widetilde \psi} [\ov {\widetilde v}_4, {\widetilde u}_4, k_4 ]  \right]&=-2 \alpha'\, g_{Dp_a} \, g_{Dp_b} \left( K + \widetilde{K}  \right) \nn \\
& \hspace{-20mm}\times  \Tr\left[T^{a}  T^{ab} T^{b} T^{ba} \right]  \frac{T}{S} B[T, U] ~,
\end{align}
with $K$ and $\widetilde{K} $ again given by \eqref{eq. K} and \eqref{eq. tilde K}, respectively.
 
Note that in the limit $\theta^1_{ab} \rightarrow 0$, thus in the limit in which the fermion $\widetilde{\psi}$ becomes massless, we get the exact same results as for the scattering amplitude of two gauge bosons with two massless fermions as computed in \cite{Lust:2008qc} and summarized in appendix \ref{AAmasslessmassless}.  

After a few simple manipulations of  the traces the amplitudes take the form
\begin{align}
&{\cal M}\left[ A^{a_1} [{\epsilon_1}, k_1], A^{a_2}[{\epsilon_2}, k_2], \widetilde{\psi}[{\widetilde v}_3, \ov {\widetilde v}_3, k_3 ], \ov {\widetilde \psi} [\ov {\widetilde v}_4, {\widetilde u}_4, k_4 ]  \right]=2 \alpha' g^2_{Dp_a} \left( K + \widetilde{K}  \right) \nn
 \\ &\hspace{30mm}
\times \frac{1}{U} \left\{  \left[T^{a_1} T^{a_2}\right]^{\alpha_3}_{\alpha_4} \delta^{\beta_4}_{\beta_3} \frac{T}{S} {\widehat V}_T +  \left[T^{a_2} T^{a_1}\right]^{\alpha_3}_{\alpha_4}  \delta^{\beta_4}_{\beta_3} \frac{U}{S} {\widehat V}_U
  \right\}
  \label{eq amplitude result 1}
  \\
&{\cal M}\left[ A^{a} [{\epsilon_1}, k_1], A^{b}[{\epsilon_2}, k_2], \widetilde{\psi}[{\widetilde v}_3, \ov {\widetilde v}_3, k_3 ], \ov {\widetilde \psi} [\ov {\widetilde v}_4, {\widetilde u}_4, k_4 ]  \right]=2 \alpha' g_{Dp_a} g_{Dp_b} \left( K + \widetilde{K}  \right) \nn
\\ &\hspace{30mm}
\times \frac{1}{U} \left[T^{a}\right]^{\alpha_3}_{\alpha_4} \left[T^{b}\right]^{\beta_4}_{\beta_3}  {\widehat V}_S
  \label{eq amplitude result 2}
\end{align}
where we introduced the \emph{generalized} Veneziano formfactor given by
\begin{align}
{\widehat V}_S =\frac{T~ U}{ T+U} B[T, U]~,~~~~~
{\widehat V}_T=\frac{S~ U}{ S+U} B[S, U] ~,~~~~~ {\widehat V}_U=\frac{S~ T}{ S+T} B[S, T] \,\,.
\end{align} 
In the following section we will take the results \eqref{eq amplitude result 1} and \eqref{eq amplitude result 2} square them and sum over all polarization and gauge configurations. Eventually we will apply our findings to the quark sector of a SM D-brane realization.

\section{Squared amplitudes \label{sec cross section}}

In this chapter we compute the squared amplitudes summed over helicities and spins, respectively, as well as over the gauge indices of initial and final particles involved in the scattering process. Eventually we will average over all the initial gauge as well as polarization configurations. 

First let us focus on the kinematics of the cross section, thus we sum over all spins and helicities of the massive fermions and gauge bosons. This involves the computation of the following expression 
\begin{align}
\sum_{\substack{h_1, h'_1 \\ h_2, h'_2}} \sum_{\substack{s_3, s'_3 \\ s_4, s'_4}}   \left( K + \widetilde{K}  \right) \,  \left( K + \widetilde{K}  \right)^{\dagger}\,\,,
\label{eq square kinematics}
\end{align}
where the $h_i$ indicate the helicities of the massless gauge bosons and the $s_i$ denote the spin polarizations of the massive fermions. Using the following completeness relations 
\begin{align}
&\hspace{30mm} \sum_{h_1, h'_1} \e_1^{*\mu} \e_1^\nu = - g^{\mu\nu} ~~~~ \\
& \sum_{s,s'} \ov {\widetilde v}^{\dot a}(k) {\widetilde v}^a(k) = -k^\mu \ov \sigma_{\mu}^{\dot a a} ~~~~ 
      \sum_{s, s'} {\widetilde u}_{a}(k) \ov {\widetilde u}_{\dot a}(k) = -k_\mu \sigma^{\mu}_{a \dot a} ~~~~ \\
& \sum_{s, s'} {\widetilde v}^{a}(k) {\widetilde u}_{b}(k) = m \delta^a_b ~~~~~~~~ \,
      \sum_{s, s'} \ov {\widetilde v}^{\dot a}(k) \ov {\widetilde u}_{\dot b}(k) = m \delta^{\dot a}_{\dot b} ~~~~ 
\end{align}
for the gauge bosons as well as the massive fermions one obtains applying various trace identities of sigma matrices displayed in appendix \ref{sigmology}
for \eqref{eq square kinematics} 
\begin{align}
\sum_{\substack{h_1, h'_1 \\ h_2, h'_2}} \sum_{\substack{s_3, s'_3 \\ s_4, s'_4}}   \left| K + \widetilde{K}  \right|^2 
=2\, \frac{U^2}{\alpha'^2} ~{\cal F}\Big[S,T,U,\sqrt{\theta^1_{ab}/\a'}\Big]\,\,,
\end{align}
where we defined for the sake of clarity
\begin{align}
{\cal F}[S,T,U,m] = \frac{T}{U} + \frac{U}{T} + 4 \a' m^2 \left( \frac{1}{T} + \frac{1}{U}\right) - 4 \a'^2 m^4 \left( \frac{1}{T} + \frac{1}{U}\right)^2\,\,.
\end{align}

The resulting squared amplitudes before summation over the gauge indices then take the form
\bea
&&\left|{\cal M}\left[ A^{a_1} [{\epsilon_1}, k_1], A^{a_2}[{\epsilon_2}, k_2], \widetilde{\psi}[{\widetilde v}_3, \ov {\widetilde v}_3, k_3 ], \ov {\widetilde \psi} [\ov {\widetilde v}_4, {\widetilde u}_4, k_4 ]  \right]\right|^2 
= 8 g_a^4 \frac{1}{S^2} {\cal F}[S,T,U,m]  ~~~~~~~~
~~~\label{eq. final Result without gauge summation 1}\\
&&~~~~~~~~~~~~~~~~~~~~~~~~~~~~~~~~\times N_b \Big\{ Tr[T^{a_1}T^{a_1}T^{a_2}T^{a_2}] (T {\widehat V}_T+U {\widehat V}_U)^2 -\frac{1}{2}f_{1,2,n}^2 U{\widehat V}_U T {\widehat V}_T \Big\}
\nn\\
&&\left|{\cal M}\left[ A^{a} [{\epsilon_1}, k_1], A^{b}[{\epsilon_2}, k_2], \widetilde{\psi}[{\widetilde v}_3, \ov {\widetilde v}_3, k_3 ], \ov {\widetilde \psi} [\ov {\widetilde v}_4, {\widetilde u}_4, k_4 ]  \right]\right|^2
=  8 g_a^2 g_b^2 {\cal F}[S,T,U,m]~~~~~
\label{eq. final Result without gauge summation 2}\nn\\
&&~~~~~~~~~~~~~~~~~~~~~~~~~~~~~~~~\times \sum_{ab} Tr[T^aT^a] Tr[T^bT^b] ~{\widehat V}_S^2\,\,,
\eea
where $f^{abc}$ denote the totally antisymmetric structure constants, $\left[T^{a}, T^b \right]= i f^{abc} T^c$. 

In the low energy limit $\alpha' \rightarrow 0$, in case of abelian gauge bosons we expect that \eqref{eq. final Result without gauge summation 1}  reduces to the Klein-Nishina formula, describing the Compton cross section. To be precise we have to take the limit $\alpha' \rightarrow 0$, while keeping $m_{\widetilde{\psi}}$ fixed, since otherwise the particle 
$\widetilde{\psi}$ becomes infinitely heavy and should be integrated out, and thus would not be part of the low energy effective action. Performing the limit gives   
for  the generalized Veneziano formfactors 
\begin{align}
&\widehat{V}_{S} = 1 - \alpha'^2 \, \frac{\pi^2}{6} \widehat{T} \widehat{U} + ...  \qquad \widehat{V}_{T} = 1 - \alpha'^2 \, \frac{\pi^2}{6}  \widehat{S}  \widehat{U}  + ...  \\
& \hspace{20mm} \widehat{V}_{U} = 1 - \alpha'^2 \, \frac{\pi^2}{6}  \widehat{S} \widehat{T}  + ...  \,\,.
\label{eq low energy limit form factors}
\end{align}
where $ \widehat{S}$, $\widehat{T}$, and  $\widehat{U}$ are the modified Mandelstam variables divided by $\alpha'$. Plugging \eqref{eq low energy limit form factors} into
 \eqref{eq. final Result without gauge summation 1} gives 
\begin{align}\label{eq. expansion=klein nishina formula}
& \lim_{\alpha' \rightarrow 0} \left|{\cal M}\left[ A^{a_1} [{\epsilon_1}, k_1], A^{a_2}[{\epsilon_2}, k_2], \widetilde{\psi}[{\widetilde v}_3, \ov {\widetilde v}_3, k_3 ], \ov {\widetilde \psi} [\ov {\widetilde v}_4, {\widetilde u}_4, k_4 ]  \right]\right|^2\\
&~~~~~~~~~~~~~~~= 8  g_a^4  \left( \frac{T}{U}+\frac{U}{T} 
+ 4 {\theta^1_{ab}} \left(\frac{1}{T}+\frac{1}{U}\right)  
- 4 {\theta^1_{ab}}^2 \left(\frac{1}{T}+\frac{1}{U}\right)^2\right) + {\cal O} [\alpha'] \,\,. \nn
\end{align}
which indeed coincides with the Klein-Nishina formula \cite{Peskin:1995ev} \footnote{The Klein-Nishina formula is usually given in terms of two incoming and two outgoing particles. Here instead we use all four momenta as incoming. In addition the authors of \cite{Peskin:1995ev} use a different metric as well as a different convention for the Mandelstam variables, which however can be easily accounted for. Moreover, in contrast to the result presented in \cite{Peskin:1995ev} we do not average over the initial polarizations.}.

Concerning the summation over the gauge indices we use the fact that the quadratic Casimir of the $SU(N)$ symmetry is
 \begin{align}
 \sum_{a}T^{a} T^{a}  = \frac{N^2-1}{2 N}    \mathbb{I}_N
 \label{eq casimir}
 \end{align}
as well as the relation
\begin{align}
\sum_{a_1, a_2,  n} f^{a_1 a_2  n} f^{a_1 a_2  n} = N(N^2-1)
\label{eq add trace condition}
\end{align}
to manipulate \eqref{eq. final Result without gauge summation 1} and \eqref{eq. final Result without gauge summation 2} to
\bea
&&\left|{\cal M}\left[ A^a A^a \rightarrow \widetilde{\psi}_{ab} \ov{\widetilde{\psi}}_{ab} ]  \right]\right|^2 = \frac{8 g_a^4}{S^2} 
{\cal F}[S,T,U,m_{\widetilde{\psi}_{ab}}]
\label{eq final result without averaging 111}\\
&&\hspace{50mm} \times  \frac{ N_b \left(N^2_a-1\right)^2}{4N_a}\left\{  (T {\widehat V}_T+U {\widehat V}_U)^2 -\frac{2 N^2_a }{N^2_a-1} U{\widehat V}_U T {\widehat V}_T \right\}
\nn\\
&&\left|{\cal M}\left[ A^a A^b \rightarrow \widetilde{\psi}_{ab} \ov{\widetilde{\psi}}_{ab} ]  \right]\right|^2 = \frac{8 g_a^2g_b^2}{S^2} {\cal F}[S,T,U,m_{\widetilde{\psi}_{ab}}] \frac{\left(N^2_a-1\right) \left(N^2_b-1\right) }{4} ~{\widehat V}_S^2\,\,.
\label{eq final result without averaging 222}
\eea
Below we will apply these formulae to the quark sector of a Standard Model D-brane realization.

\subsection*{Connection to the Standard Model}

\begin{figure}[h]
\begin{center}
\epsfig{file=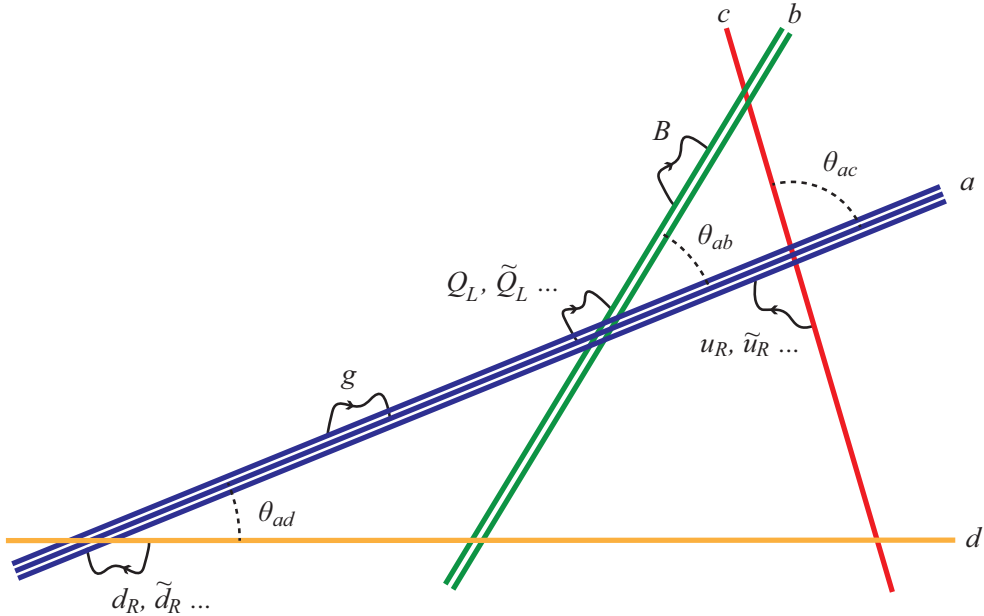,width=130mm}
\caption{Quarks and their stringy excitations of a SM D-brane realization.
}\label{fig.Quark D-brane}
\end{center}
\end{figure}

In the following we apply the previously carried out computation to D-brane realization of the a Standard Model. We will focus on the stringy excitations of the quark sector, but an analogous analysis can be carried out for the stringy excitations of the leptonic sector as well. Recall that the left-handed quarks $Q_L$ and its stringy excitations $\widetilde{Q}_L$ arise from the intersection of color D-brane stack and the $SU(2)_L$ D-brane stack. On the other hand the right-handed quarks $d_R$ and its stringy excitations $\widetilde{d}_R$, as well as $u_R$ and its stringy excitation $\widetilde{u}_R$ are localized at the intersection of the color D-brane stack and a $U(1)$ D-brane stack, which we will call $c$ and $d$, respectively \footnote{Here we assume that the quarks are always realized as bi-fundamentals and moreover, that all three families arise from the same D-brane stack intersection. One can easily generalise  this scenario to setups in which the different families are differently charged with 
respect to D-brane gauge symmetries. For original work on local D-brane configurations, see \cite{Antoniadis:2000ena,Aldazabal:2000sa}. For a systematic analysis of local D-brane configurations, see \cite{Gmeiner:2005vz,Anastasopoulos:2006da,Cvetic:2009yh,Cvetic:2010mm}.} (see figure \ref{fig.Quark D-brane}).

Note that each D-brane intersection will give rise to different intersection angles and thus stringy excitations localized at different intersections will have different masses. Let us start by considering gluon fusion into massive stringy excitations of the left-quarks, namely $\widetilde{Q}_L$ and $ \ov{\widetilde{Q}}_L$. The latter are localized between the color D-brane stack $a$ and the $SU(2)_L$ D-brane stack $b$, thus between two non-abelian D-brane stacks.

Using \eqref{eq final result without averaging 111}, and averaging over all initial states which implies dividing by $2(N_a^2-1)$ for each gauge boson due to the two helicities and the dimension of the adjoint representation of $SU(N_a)$, we get
\begin{align}
\label{eq gg --> qq}
&\left|{\cal M}\left[ g g \rightarrow \widetilde{Q}_L \ov{\widetilde{Q}}_L ]  \right]\right|^2
= g^4 \frac{1}{S^2} 
{\cal F}[S,T,U,m_{\widetilde{Q}_L}] ~ \frac{1}{3}\left\{  (T {\widehat V}_T+U {\widehat V}_U)^2 -\frac{9}{4} U{\widehat V}_U T {\widehat V}_T \right\}\,\,,
\end{align}
where $\alpha' m^2_{\widetilde{Q}_L}= \theta_{ab} $ \footnote{The mass of the lightest stringy excitation depends on the smallest intersection angle. In the explicit computation we assume the smallest of the three intersection angles to be $\theta^1_{ab}$. As pointed out before the analysis is independent of the choice which of the three intersection angles is the smallest.}, $g$ denoting the color gauge coupling, and $N_a=3$ and $N_b=2$ due to the fact that ${\widetilde Q}_L$ is stretched between an $SU(3)$ and $SU(2)$ D-brane stack. 

On the other hand the gluon scattering into the stringy excitations of the right-handed quarks, $\widetilde{u}_R$ and $\widetilde{d}_R$ take the form
\begin{align}
\label{eq gg --> dd}
&\left|{\cal M}\left[ g g \rightarrow \widetilde{d}_R \ov{\widetilde{d}}_R ]  \right]\right|^2
= g^4 \frac{1}{S^2} 
{\cal F}[S,T,U,m_{\widetilde{d}_R}]~  \frac{1}{6}\left\{  (T {\widehat V}_T+U {\widehat V}_U)^2 -\frac{9}{4} U{\widehat V}_U T {\widehat V}_T \right\}\\
\label{eq gg --> uu}
&\left|{\cal M}\left[ g g \rightarrow \widetilde{u}_R \ov{\widetilde{u}}_R ]  \right]\right|^2
=  g^4 \frac{1}{S^2} 
{\cal F}[S,T,U,m_{\widetilde{u}_R}]~\frac{1}{6}\left\{  (T {\widehat V}_T+U {\widehat V}_U)^2 -\frac{9}{4} U{\widehat V}_U T {\widehat V}_T \right\}
\end{align}
with $\alpha' m^2_{\widetilde{d}_R} = \theta_{ac}$ and $\alpha' m^2_{\widetilde{u}_R} = \theta_{ad}$, respectively and the appropriate modifications in the definitions of $S$, $T$ and $U$ compared to \eqref{eq def of S T U}. Note that in contrast to the scattering into stringy excitations of the left-handed quarks $\widetilde{Q}_L$, where the spectator  D-brane was a $U(2)$ D-brane stack, in \eqref{eq gg --> dd} as well as  \eqref{eq gg --> uu} the spectator D-brane stack is a $U(1)$ stack, explaining the factor of $2$ difference, compared to \eqref{eq gg --> qq}. 

From \eqref{eq final result without averaging 111} and  \eqref{eq final result without averaging 222} we can also deduce other processes by using the crossing relations. Those include the scattering of a gluon onto a stringy excitation of the quarks as well as the annihilations of stringy excitations of the quarks into gluons. The results of those processes are summarized in table \ref{table processes}. Note that the definitions of $S$, $T$ and $U$ crucially depend on the constituents of the process and generically are different for each scattering process\footnote{In this work we use the conventions of all momenta being incoming.}. Again we averaged over all initial polarization and gauge configurations.

Finally, we also display the processes in which a $B$ boson, the gauge boson living on the $SU(2)$ D-brane stack, is involved. The squared amplitude can be derived from  \eqref{eq final result without averaging 222} by averaging over the $2(N_a^2-1)$ and  $2(N_b^2-1)$ initial configurations of $g$ and $B$, respectively. For $N_a=3$ and $N_b=2$ one obtains
\begin{align} 
\label{eq g B -> QQ}
&\left|{\cal M}\left[ g B  \rightarrow  \widetilde{Q}_L  \ov {\widetilde Q}_L \right]\right|^2 
=  \frac{1}{2} g^2 g_b^2  
{\cal F}[S,T,U,m_{\widetilde{Q}_L}]{\widehat V}_S^2\,\,,
\end{align}
where $g$ again denotes the color gauge coupling and $g_B$ is the gauge coupling of the $B$-boson gauge group. From \eqref{eq g B -> QQ} we can determine all other processes using the crossing relations in an analogous fashion as done above. We display the processes in table \ref{table processes} where we again average over all the initial polarization and gauge configurations.

\begin{table}[h] \centering
\begin{tabular}{| c | c |l}
\hline
process & $\bigg. |{\cal M}|^2$ \\ \hline \hline
$~~~~\Bigg. g\, g \rightarrow \widetilde{Q}_L \, \ov {\widetilde Q}_L   ~~~~$ & $~~~~g^4 \frac{1 }{S^2} {\cal F}[S, T, U, m_{\widetilde{Q}_L}]  \,\,  \frac{1}{3}\left\{  (T {\widehat V}_T+U {\widehat V}_U)^2 -\frac{9}{4} U{\widehat V}_U T {\widehat V}_T \right\}  ~~~~$ \\
 \hline
$\Bigg. g\, g \rightarrow \widetilde{d}_R \, \ov {\widetilde d}_R   $ & $g^4 \frac{1}{S^2} {\cal F}[S, T, U, m_{\widetilde{d}_R}]  \,\,  \frac{1}{6}\left\{  (T {\widehat V}_T+U {\widehat V}_U)^2 -\frac{9}{4} U{\widehat V}_U T {\widehat V}_T \right\}  $
\\
 \hline
$\Bigg. g \,g \rightarrow \widetilde{u}_R\,  \ov {\widetilde u}_R   $ & $g^4 \frac{1 }{S^2} {\cal F}[S, T, U, m_{\widetilde{u}_R}]  \,\,  \frac{1}{6}\left\{  (T {\widehat V}_T+U {\widehat V}_U)^2 -\frac{9}{4} U{\widehat V}_U T {\widehat V}_T \right\}  $ \\ \hline \hline
$\Bigg. g \,  \widetilde{Q}_L \rightarrow g \,\ov {\widetilde Q}_L   $ & $g^4 \frac{1}{T^2} {\cal F}[T, S, U, m_{\widetilde{Q}_L}]  \,\,  \frac{4}{9}\left\{  (S {\widehat V}_S+U {\widehat V}_U)^2 -\frac{9}{4} U{\widehat V}_U S {\widehat V}_S \right\} $\\ \hline
$\Bigg. g   \,\widetilde{d}_R \rightarrow g \,\ov {\widetilde d}_R   $ & $g^4 \frac{1}{T^2} {\cal F}[T, S, U, m_{\widetilde{d}_R}]  \,\,  \frac{4}{9}\left\{  (S {\widehat V}_S+U {\widehat V}_U)^2 -\frac{9}{4} U{\widehat V}_U S {\widehat V}_S \right\} $ \\
\hline
$\Bigg. g   \,\widetilde{u}_R \rightarrow g \,\ov {\widetilde u}_R   $ & $g^4 \frac{1 }{T^2} {\cal F}[T, S, U, m_{\widetilde{u}_R}]  \,\,  \frac{4}{9}\left\{  (S {\widehat V}_S+U {\widehat V}_U)^2 -\frac{9}{4} U{\widehat V}_U S {\widehat V}_S \right\} $ \\ \hline
$\Bigg. g \,  \widetilde{Q}_L \rightarrow B \,\ov {\widetilde Q}_L   $ & $  g^2 g^2_B  {\cal F}[T, S, U, m_{\widetilde{Q}_L}]  \,\,    \frac{1}{4}    {\widehat V}_T^2$\\ \hline \hline
$\Bigg.  \widetilde{Q}_L \, \ov {\widetilde Q}_L   \rightarrow g \, g $ & $g^4 \frac{1}{S^2} {\cal F}[S, T, U, m_{\widetilde{Q}_L}]  \,\,  \frac{16}{27}\left\{  (T {\widehat V}_T+U {\widehat V}_U)^2 -\frac{9}{4} U{\widehat V}_U T {\widehat V}_T \right\}  $ \\ \hline
$\Bigg.  \widetilde{d}_R \, \ov {\widetilde d}_R   \rightarrow g \, g $ & $g^4 \frac{1}{S^2} {\cal F}[S, T, U, m_{\widetilde{d}_R}]  \,\,  \frac{32}{27}\left\{  (T {\widehat V}_T+U {\widehat V}_U)^2 -\frac{9}{4} U{\widehat V}_U T {\widehat V}_T \right\}  $ \\ \hline
$\Bigg.  \widetilde{u}_R \, \ov {\widetilde u}_R   \rightarrow g \, g $ & $g^4 \frac{1 }{S^2} {\cal F}[S, T, U, m_{\widetilde{u}_R}]  \,\,  \frac{32}{27}\left\{  (T {\widehat V}_T+U {\widehat V}_U)^2 -\frac{9}{4} U{\widehat V}_U T {\widehat V}_T \right\}  $ \\ \hline
$\Bigg.  \widetilde{Q}_L \, \ov {\widetilde Q}_L   \rightarrow g \, B $ & $g^2 g^2_B  {\cal F}[S, T, U, m_{\widetilde{Q}_L}]  \,\,    \frac{1}{3}    {\widehat V}_S^2$ \\ \hline

\end{tabular}\nn

\caption{\small {Processes involving two gauge bosons and stringy excitations of quarks. }} 
\label{table processes}
\end{table}

\section{Conclusions}

In this work we investigate stringy excitations of states localized at the intersection of two D-brane stacks.  We argue that the mass of those so called light stringy states can be as low as the TeV scale, in compactifications which lead to a low string scale such as large volume scenarios. It should be stressed that those light stringy states can be significantly lighter than the Regge states of the gauge bosons discussed in \cite{Lust:2008qc,Anchordoqui:2009mm,Lust:2009pz,Anchordoqui:2009ja}. Moreover, in contrast to Kaluza Klein theories, where to each massless (SM-) particle there is a tower of massive states with same mass spacing, here for each massless particle there exists a tower of massive states, however all those particle towers  come with different mass spacing \footnote{This holds only true for massless particles, that are localized at intersections of different D-brane stacks. Thus for different families, that appear at the intersections of the same two D-brane stacks, the tower of stringy 
excitations has the same mass spacing.}. This is due to the 
fact that the mass gaps depend on one hand on the string scale $M_s$ but also on the intersection angle, which generically is different for different particles.  

We compute the disk string scattering amplitude containing two gauge bosons and two of fermionic light stringy states. As a preparatory but crucial step we determine the explicit vertex operators of light stringy states extending previous investigations \cite{Anastasopoulos:2011hj}. Eventually we sum over all polarizations and color configurations of the initial and final states. In the limit of $\theta_{ab} \rightarrow 0$, i.e. in the limit in which those light stringy states become massless, we recover the results 
of \cite{Lust:2008qc}, which analyzed the scattering of two gauge bosons onto two massless fermions. Moreover, we show assuming abelian gauge bosons that in the low energy limit our squared amplitude coincides with the Klein-Nishina formula.

Finally we apply our computation to the scattering of two SM gauge bosons onto two stringy excitations of the quark sector. The squared amplitudes summed over all final polarization and gauge configurations as well as averaged over all initial ones are summarized in table \ref{table processes}. 

\newpage

\section*{Acknowledgements}

We thank I. Antoniadis, M. Bianchi, R. Boels, W. Z. Feng, M. Goodsell, T. Hansen, C. Horst, E. Kiritsis, I. Pesando, O. Schlotterer, S. Stieberger and W. Waltenberger for useful discussions. P.~A. is supported by the Austrian Science Fund (FWF) program M 1428-N27.
R.~R. was partly supported by the German Science Foundation (DFG) under the Collaborative Research Center (SFB) 676 ``Particles, Strings and the Early Universe".
Robert Richter acknowledges the support from the FP7 Marie Curie Actions of the European Commission, via the Intra-European Fellowships (project number: 328170). The authors would like to thank Bonn University, CERN, Ecole Polytechnique, University of ``Tor Vergata'', Torino University for hospitality during parts of this work.

\newpage
\appendix

\section{OPE's and correlators \label{app OPE}}

The OPE's of the regular and excited boson twist fields take the form \cite{Anastasopoulos:2011gn, Anastasopoulos:2013sta}\footnote{Note the different prefactors compared to the OPE's displayed in \cite{Anastasopoulos:2011gn, Anastasopoulos:2013sta}, e.g. for the OPE  $\partial \ov Z \, \widetilde\tau$. The appearance of the angle dependent pre facto can be easily derived as demonstrated below. Let us recall the mode expansion of $\partial Z $ and $\partial \ov Z$ that take the form
 \begin{align}
\partial Z (z) = \sum_{n}
\alpha_{n-\theta} \,\,z^{-n+\theta-1} \qquad \qquad  \partial \ov Z (z) = \sum_{n}
\alpha_{n+\theta} \,\,z^{-n-\theta-1}\,\,,
\end{align}
where the only non-vanishing commutators  are
$[\alpha_{n \pm \theta}, \alpha_{m \mp \theta} ] =  (m \pm \theta)$.
For positive angles the ground state is given by 
\begin{equation}
\begin{aligned}
\alpha_{m-\theta} |\,\theta\,\rangle &=0   \qquad m \geq 1 \qquad
\qquad
 \alpha_{m+\theta} |\,\theta\,\rangle &=0 \qquad m \geq 0\,\,,
 \end{aligned}
\end{equation}
where the ground state $|\,\theta\,\rangle $ is identified with the bosonic twist field $\sigma$.
Let us now act with $\partial \ov Z$ on the ground state $|\,\theta\,\rangle \sim \sigma(0) $
\begin{align}
 \lim_{z \rightarrow 0} \partial \ov Z(z)    |\,\theta\,\rangle =   \sum_{n}
\alpha_{n+\theta} \,\,z^{-n-\theta-1}  |\,\theta\,\rangle\,\, \sim z^{-\theta} \alpha_{-1+\theta}  |\,\theta\,\rangle\,\,.
\end{align}
Thus $\widetilde{\tau}$ is identified with the state $\alpha_{-1+\theta}  |\,\theta\,\rangle$, reproducing the above OPE's.  
Acting with the $\partial Z$ on $\alpha_{-1+\theta}  |\,\theta\,\rangle \sim \widetilde{\tau} (0)$ one obtains
\begin{align}
\rightarrow  \lim_{z \rightarrow 0} \partial  Z(z) \,\,\alpha_{-1+\theta}   |\,\theta\,\rangle 
=  \sum_{n}
\alpha_{n-\theta} \,\,z^{-n+\theta-1}   \,\,\alpha_{-1+\theta}  |\,\theta\,\rangle\,\,
\sim 
(1-\theta)  z^{-2+\theta}  |\,\theta\,\rangle\,\,,
\end{align}
where we used the commutation relations. 
Thus we conclude 
\begin{align}
\partial Z(z) \widetilde{\tau}_{\theta}(w) \sim (1-\theta)(z-w)^{-2+\theta} \sigma_{\theta} \,\,.
\end{align}
Analogously one can derive all other OPE's involving the bosonic twist fields.

}
\bea
\ba{lllll}
& \partial Z(z) \, \sigma^+_{a} (w) \sim   \sqrt{2 \alpha'}  \,(z-w)^{a-1} \, \tau^+_{a} (w)  & &  \partial \ov Z (z) \, \sigma^+_{a} (w) \sim  \sqrt{2 \alpha'}  (z-w)^{-a} \widetilde{\tau}^+_{a} (w) \\ 
& \partial Z (z) \, \tau^+_{a} (w) \sim \sqrt{2 \alpha'} (z-w)^{a-1} \omega^+_{a} (w)  &&  \partial \ov Z(z) \, \tau^+_{a} (w) \sim\sqrt{2 \alpha'}  a(z-w)^{-a-1} \sigma^+_{a} (w)  \\
& \partial Z (z) \, \widetilde{\tau}^+_{a} (w) \sim \sqrt{2 \alpha'}  (1-a) (z-w)^{-2+a} \sigma^+_{a} (w)  && \partial \ov Z (z) \, \widetilde{\tau}^+_{a} (w) \sim \sqrt{2 \alpha'} (z-w)^{-a} \widetilde{\omega}^+_{a} (w)
\nn 
\ea
\eea
where the conformal dimensions of the respective twist fields are
\bea
\sigma^+_{a}  ~:~   \frac{1}{2}a(1-a)  ~,~~~~~ \tau^+_{a}     ~:~   \frac{1}{2}a(3-a)   ~,~~~~~ 
 \widetilde{\tau}^+_{a}     ~:~   \frac{1}{2}(a+2)(1-a)~~~  .
\eea
As pointed out in \cite{Anastasopoulos:2011gn} bosonic anti-twist fields can be written as bosonic twist fields with the angle $a$ replaced by $1-a$, for instance $\sigma^{-}_{a} (z) = \sigma^+_{1-a} (z)$.

In addition throughout this work we need the following OPE's
\begin{align}
 e^{a\phi(z)} e^{b\phi(w)} &\sim \frac{e^{(a+b)\phi(w)}}{(z-w)^{ab}}  
\qquad \qquad  \hspace{10.2mm}
e^{aH(z)} e^{bH(w)} \sim \frac{ e^{(a+b)H(w)}  }{(z-w)^{ab}}
  \\
 \partial^\mu X(z) e^{ikX(w)} &\sim -\frac{2i\alpha' k^\mu}{z-w} e^{ikX(w)}  \qquad \qquad
 \psi^\mu(z) \psi^\nu(w) \sim \frac{\eta^{\mu \nu}} {(z-w)}     \\
  \psi^\mu(z) S_{a} (w) &\sim \frac{1}{\sqrt{2}} \frac{\sigma^\mu_{a \dot a} S^{\dot a}(w)}{(z-w)^{1/2}}  \qquad \qquad \hspace{2.3mm}
 \psi_\mu(z) S^{\dot a} (w) \sim \frac{1}{\sqrt{2}} \frac{\bar \sigma_\mu^{\dot a a} S_a(w)}{(z-w)^{1/2}}  \,\,.
\end{align}
For the string amplitude computation we apply the correlators \footnote{For twist correlators involving higher excited bosonic twist fields, see \cite{David:2000yn,Lust:2004cx, Conlon:2011jq, Pesando:2011ce,Pesando:2012cx, Anastasopoulos:2013sta, Pesando:2014owa}.}
\begin{align} \nn
&\langle \sigma_a(x_1)  \sigma_{1-a}(x_2)\rangle = x_{12}^{-a(1-a)}\qquad \qquad 
\langle \tau_a(x_1)  \widetilde\tau_{1-a}(x_2) \rangle =  a x_{12}^{-a(3-a)}\\
& \hspace{3cm} \left\langle e^{i a H (x_1)} e^{-i a H (x_2)} \right\rangle = x^{a^2}_{12} 
\end{align}
as well as 
\begin{align}
 \langle e^{-\varphi(x_1)} e^{-\frac{\varphi(x_2)}{2}} e^{-\frac{\varphi(x_3)}{2}}  \rangle 
&= x_{12}^{-1/2} x_{13}^{-1/2} x_{23}^{-1/4}\\
 \langle e^{ik_1X(x_1)} e^{ik_2X(x_2)} e^{ik_3X(x_3)}  e^{ik_4X(x_4)} \rangle 
&= \prod_{i\neq j} x_{ij}^{2\alpha' k_i k_j}  
\\
 \langle 
\partial X^\nu(x_1)
e^{ik_1X(x_1)} e^{ik_2X(x_2)} e^{ik_3X(x_3)}  e^{ik_4X(x_4)} \rangle &=  \prod_{i\neq j} x_{ij}^{2\alpha' k_i k_j}   \sum_{n\neq 1} \frac{  -2i\a' k_n^{\nu}}{x_{1i}} \\
\langle \psi^\mu(x_3) S_a(x_4) S_{\dot a}(x_5) \rangle &= \frac{1}{\sqrt{2}} 
\frac{\sigma^\mu_{a\dot a}}{(x_{34} x_{35})^{1/2}}\\
 \langle \psi_\mu(x_3) S^{\dot a}(x_4) S^{a}(x_5) \rangle &= \frac{1}{\sqrt{2}} 
\frac{\bar \sigma_\mu^{\dot a a}}{(x_{34} x_{35})^{1/2}}\\
\langle \psi^\mu(x_1)  \psi^\rho(x_2)  \psi^\nu(x_3) S_a(x_4) S_{\dot a}(x_5) \rangle &= \frac{1}{\sqrt{2}} 
(x_{14} x_{15} x_{24} x_{25} x_{34} x_{35})^{-1/2}\\
&\hspace{-6cm}\times \left(\frac{z_{45}}{2}(\sigma^\mu\bar \sigma^\rho\sigma^\nu)_{a\dot a} 
+ g^{\mu\rho}\sigma^\nu_{a\dot a}\frac{x_{14}x_{25}}{x_{12}}
- g^{\mu\nu}\sigma^\rho_{a\dot a}\frac{x_{14}x_{35}}{x_{13}}
+ g^{\nu\rho}\sigma^\mu_{a\dot a}\frac{x_{24}x_{35}}{x_{23}}\right)\nn
\end{align}

\section{State - Vertex operator dictionary \label{app dictionary} }

In table \ref{table R-sector} we display the dictionary between state and their corresponding vertex operators in the Ramond sector, which we apply in chapter \ref{sec vertex operator} to determine the vertex operators of the massless fermion $\psi$ and the fermionic light stringy state $\widetilde{\psi}$. For more details, see \cite{Anastasopoulos:2011hj}
\begin{table}[h] \centering
\begin{tabular}{ l  l  l  l }
\multicolumn{2}{l}{Positive angles}&
\multicolumn{2}{l}{Negative angles}
\\
state & vertex operator ~~~~~~~ &state & vertex operator ~~~~~~~\\
$ | \, \theta \, \rangle_R $ & 
$  e^{i \left(\theta-\frac{1}{2}\right) H(z)} \sigma^+_{\theta}(z) $ & 
$ | \, \theta \, \rangle_R $ & 
$  e^{i \left(\frac{1}{2}+\theta\right) H(z) } \sigma^-_{-\theta}(z) $\\ 
$ \alpha_{-\theta} | \, \theta\, \rangle_R $ & 
$  e^{i \left(\theta-\frac{1}{2}\right) H(z) }\tau^+_{\theta}(z) $ & 
$ \alpha_{\theta} | \, \theta\, \rangle_R $ & 
$  e^{i \left(\frac{1}{2}+\theta\right) H(z) }\widetilde\tau^-_{-\theta}(z) $ \\ 
$ \psi_{- \theta}| \, \theta \, \rangle_R  $ &  
$  e^{i \left(\theta+\frac{1}{2}\right)H(z)} \sigma^+_{\theta}(z) $ & 
$ \psi_{ \theta}| \, \theta \, \rangle_R $ & 
$  e^{i \left(\theta +\frac{1}{2}\right)H(z)} \sigma^-_{-\theta}(z) $ \\ 
$ \alpha_{-\theta}\, \psi_{-\theta} | \, \theta\, \rangle_R $ &  
$  e^{i \left(\theta +\frac{1}{2}\right) H(z) }\tau^+_{\theta}(z) $ &
$ \alpha_{\theta}\, \psi_{\theta} | \, \theta\, \rangle_R $ & 
$ e^{i \left(\theta +\frac{1}{2}\right)H(z) }\widetilde\tau^-_{-\theta}(z) $ \\ 
\end{tabular}
\caption{\small {Excitations in the R-sector and their corresponding vertex operator part.}} 
\label{table R-sector}
\end{table}

\section{Trace identities}\label{sigmology}

Here we present some trace identities used throughout the chapter \ref{sec cross section} \footnote{Note that the metric signature is given by $(-,+,+,+)$.}
\bea \nn
&&g^{\mu \nu}g_{\mu \nu} =4 ~~~~~~~ 
\qquad \qquad 
\sigma_{a\dot a}^\mu \bar \sigma^{\dot b b}_\mu = -2 \d^{b}_{a} \d^{\dot b}_{\dot a}\\ \nn
&& \sigma^\mu \bar \sigma^\nu \sigma^\rho = - g^{\mu\nu} \sigma^\rho + g^{\mu\rho} \sigma^\nu - g^{\rho\nu} \sigma^\mu - i \e^{\mu\nu\rho\k} \sigma_\k ~\\
&& \sigma^\mu \bar \sigma^\nu \sigma^\rho \bar \sigma_\m = - \sigma^\rho\bar \sigma^\nu + \sigma^\nu \bar \sigma^\rho - g^{\rho\nu} \sigma.\bar \sigma - i \e^{\mu\nu\rho\k} \sigma_\k\bar \sigma_\mu ~\\ \nn
&& Tr[\sigma^\mu \bar \sigma^\nu] =Tr[\bar \sigma^\mu \sigma^\nu] = -2 g^{\mu\nu} ~\\ \nn
&& Tr[\sigma^\mu \bar \sigma^\nu \sigma^\rho \bar \sigma^\k ]= 2 (g^{\mu\nu}  g^{\rho\k} -  g^{\mu\rho}  g^{\nu\k} + g^{\mu\k}  g^{\nu\rho}
+ i \e^{\mu\nu\rho\k}) ~. 
\eea

\section{The amplitude $\langle A^\mu \, A^\nu \, \psi \, \ov \psi \rangle  $}\label{AAmasslessmassless}

In this section, we recall the key steps in the computation of the four-point scattering amplitude of two gauge bosons into two massless chiral fermions localized at the intersection of two D6-branes \cite{Lust:2008qc}
\begin{align} \nn
{\cal M}=\int \frac{\prod^4_ {i=1} dz_i }{ V_{CKG}}
  \Big\langle V^{(0)}_A [z_1, \e_1, k_1] \,  V^{(-1)}_A [z_2, \e_2, k_2] \,V^{(-1/2)}_{{\psi}} [z_3 , v_3, k_3 ] 
 \,V^{(-1/2)}_{\ov {\psi }} [z_4, \ov v_4, k_4 ] 
\, \Big\rangle \,\,.
\end{align}
The first three vertex operators are given in \eqref{eq gauge boson -1 ghost}, \eqref{eq gauge boson 0 ghost} and  \eqref{eq vertex operator massless}, respectively, while the last one takes the form
\bea
V^{(-1/2)}_{\ov \psi} = g_{\psi}\,  [T^{ab}]^{\beta_1}_{\alpha_1} \, e^{-\varphi/2}   \,\ov v_{\dot \alpha} \,S^{\dot \alpha} \,\,\prod^2_{I=1} \sigma_{1-\theta^I_{ab}} \, e^{-i\left(\theta^I_{ab}- \frac{1}{2} \right)H_I} \, \sigma_{-\theta^3_{ab}} \, e^{-i\left(\theta^3_{ab}- \frac{1}{2} \right)H_3}   \,\, e^{ikX} \,\,.~~~~~~~
  \label{eq vertex operator bar massless}
\eea

Applying the correlators displayed in appendix \ref{app OPE} one obtains
\bea
{\cal M} &\sim& \sqrt{\alpha'} \,\widetilde{C}_{D2}\, g_{A_x}  g_{A_y}\,
g^2_{\psi}  \int \frac{\prod^4_{i=1} d x_i}{ V_{CKG}} \,\,x^{s-1}_{12}  \,  x^{t}_{13} \,  x^{u-1}_{14} x^{u-1}_{23} \,  x^{t}_{24}  \, x^{s-1}_{34}  \\
&&\times \left\{ \left(  (k_2 \cdot \epsilon_1) {\epsilon_2}_{\nu}  - ( k_1 \cdot \epsilon_2) {\epsilon_1}_{\nu} + (\epsilon_1 \cdot \epsilon_2) {k_1}_{\nu}  + \frac{x_{12}\, x_{34}}{x_{13}\, x_{24}} (k_3 \cdot \epsilon_1){\epsilon_2}_{\nu} 
\right) 
\left( {v}_3 \, \sigma^{\nu} {\ov {v}}_4 \right)  \right.~~~~ \nn \\ 
&& ~~~~~~~~~~~~~~~~~~~~~~~~~~~~~~~~~~~~~~~~~~~~~~~~ \left.+\frac{1}{2} \frac{x_{12}\, x_{34}}{x_{13}\, x_{24}} \, {\epsilon_1}_{\mu}\,  {\epsilon_2}_{\nu} \, {k_1}_{\lambda} \left( {\widetilde v}_3 \, \sigma^{\lambda} \ov \sigma^{\mu} \sigma^{\nu} \, {\ov {\widetilde v}}_4 \right)
\right\} ~.\nn
\eea
Fixing $x_1\to 0,~ x_3\to 1,~ x_4\to \infty$ yields analogously to the computation carried out in chapter \ref{sec scattering amplitude}
\begin{align}
\label{eq. amplitude massless 1}
&{\cal M}\left[ A^{a_1} [{\epsilon_1}, k_1], A^{a_2}[{\epsilon_2}, k_2], {\psi}[{ v}_3, k_3 ], \ov { \psi} [\ov { v}_4, k_4 ]  \right]=2 \alpha' g^2_{Dp_a} K  \nn
 \\ &\hspace{48mm}
\times \frac{1}{u} \left\{  \left[T^{a_1} T^{a_2}\right]^{\alpha_3}_{\alpha_4}  \delta^{\beta_4}_{\beta_3}  \frac{t}{s} {\widehat V}_t +  \left[T^{a_2} T^{a_1}\right]^{\alpha_3}_{\alpha_4}  \delta^{\beta_4}_{\beta_3}  \frac{u}{s} {\widehat V}_u
 \right\}~~~~\\
 \label{eq. amplitude massless 2}
&{\cal M}\left[ A^{a} [{\epsilon_1}, k_1], A^{b}[{\epsilon_2}, k_2], {\psi}[{ v}_3, \ov { \psi} [\ov { v}_4, k_4 ]  \right]=2 \alpha' g_{Dp_a} g_{Dp_b} K 
~\frac{1}{u} \left[T^{a}\right]^{\alpha_3}_{\alpha_4} \left[T^{b}\right]^{\beta_4}_{\beta_3}  {\widehat V}_s~\,\,,
\end{align}
where for amplitude \eqref{eq. amplitude massless 1} the two gauge bosons arise form the same D-brane stack while for \eqref{eq. amplitude massless 2} the two gauge bosons arise from two different D-brane stacks. In both amplitudes the kinematic factor is given by 
\begin{align}
K&= \left\{ (k_2 \cdot \epsilon_1) {\epsilon_2}_{\nu}  - ( k_1 \cdot \epsilon_2) {\epsilon_1}_{\nu} + (\epsilon_1 \cdot \epsilon_2) {k_1}_{\nu}  - \frac{s}{t} (k_3 \cdot \epsilon_1){\epsilon_2}_{\nu} \right\} ( { v}_3 \, \sigma^{\nu} {\ov { v}}_4 )  \nn \\
& \hspace{30mm}-\frac{1}{2} \frac{s}{t} \, {\epsilon_1}_{\mu}\,  {\epsilon_2}_{\nu} \, {k_1}_{\lambda} \left(  { v}_3 \, \sigma^{\lambda} \ov \sigma^{\mu} \sigma^{\nu} \, {\ov { v}}_4 \right) \,\,.\label{eq. Kmassless}
\end{align}
Finally, the squared amplitudes, before summation over the gauge configurations take the form
\begin{align}
&\left|{\cal M}\left[ A^{a_1} [{\epsilon_1}, k_1], A^{a_2}[{\epsilon_2}, k_2], {\psi}[{ v}_3, k_3 ], \ov { \psi} [\ov { v}_4, k_4 ]  \right]\right|^2 
\label{eq. massless final Result without gauge summation 1}\\
& ~~~~~~~~~~~~~~ = 8 g_a^4 \frac{1}{s^2} \left\{ \frac{t}{u}+\frac{u}{t} \right\} N_b \Big( Tr[T^{a_1}T^{a_1}T^{a_2}T^{a_2}] (t \widehat{V}_t+u \widehat{V}_u)^2 -\frac{1}{2}f_{1,2,n}^2 u\widehat{V}_u t \widehat{V}_t \Big)
\nn\\
&\left|{\cal M}\left[ A^{a} [{\epsilon_1}, k_1], A^{b}[{\epsilon_2}, k_2], {\psi}[{ v}_3, k_3 ], \ov { \psi} [\ov { v}_4, k_4 ]  \right]\right|^2 \nn\\
& ~~~~~~~~~~~~~~ =  8g_a^2 g_b^2 \left\{ \frac{t}{u}+\frac{u}{t}\right\} \sum_{ab} Tr[T^aT^a] Tr[T^bT^b] ~\widehat{V}_s^2\,\,.
\label{eq. massless final Result without gauge summation 2}
\end{align}

\clearpage \nocite{*}
\providecommand{\href}[2]{#2}\begingroup\raggedright\endgroup

\end{document}